# The Development of an Inquiry-based Curriculum Specifically for the Introductory Algebra-based Physics Course

Beth Thacker and Abel Diaz, Physics Department, Box 41051, Texas Tech University, Lubbock, TX 79416-1051 and Ann Marie Eligon[(a)], St Petersburg College Department of Science, Engineering and Wellness, P.O. Box 13489, St Petersburg FL 33733-3489

## Abstract

We discuss an inquiry-based curriculum that has been developed specifically for the introductory algebra-based physics course, taking into account the needs, backgrounds, learning styles and career goals of the students in that class. The course is designed to be taught in a laboratory-based, "Workshop Physics"[1] style environment, however parts of the materials can be used in other settings. As instructors we found ourselves drawing on materials developed for the calculus-based course and for other populations (materials developed for pre-service teachers, for example), parts of which were appropriate, but not a complete curriculum as we would like to teach it, developed specifically for students in the introductory algebra-based physics course. So we have modified and adapted parts of existing materials and integrated them with our own new units and our own format, creating a course aimed specifically at these students.

[1]Priscilla W. Laws, *Workshop Physics Activity Guide*, (John Wiley & Sons, Inc., New York, 1999).

[(a)] Ann Marie Eligon was at Physics Department, 1 Campus Drive, Grand Valley State University, Allendale, MI 49401 at the time of this work.

# The Development of a Curriculum Specifically for the Introductory Algebra-based Physics Course

Beth Ann Thacker and Abel Diaz, Physics Department, Box 41051, Texas Tech University, Lubbock, TX 79416-1051 and Ann Marie Eligon[a], St Petersburg College Department of Science, Engineering and Wellness, P.O. Box 13489, St Petersburg FL 33733-3489

Until recently, the introductory algebra-based physics sequence has received little attention. While individual people have made it a concern and have written materials specifically for students in the algebra-based physics class, most efforts have been local and not widely disseminated or have resulted in a text.[1] The course has not been the focus of major curricular changes: in content, instructional methods, scientific skills to be learned (modeling, mathematical modeling, etc.), or the use of the results of physics education research. We have written a curriculum, not a text, that takes the needs, backgrounds, learning styles and career goals of the students in that class into account.

As instructors of the introductory algebra-based course, we often found ourselves drawing on materials that had been developed for the calculus-based course or for other populations (materials developed for pre-service teachers, for example). Parts of these materials were appropriate, but there were parts that were not appropriate for our students that had to be dropped or modified. We wanted a set of materials designed specifically for students in the algebra-based physics course that did not have to be modified, taking their needs, learning styles, backgrounds and motivations into account. However, since there are overlaps, not so much in needs, backgrounds and learning styles, but in conceptual understanding, for example, with other student populations, it did not seem reasonable to create a whole new curriculum, starting from scratch.

We wrote and received a grant from the National Science Foundation to adapt Workshop Physics[2] materials to the algebra-based physics course at our locations, adding new units as appropriate and modifying existing materials to meet our needs.[3] We ended up creating an inquiry-based curriculum specifically for the algebra-based physics course that consists of parts modified and adapted from a number of existing materials integrated with our own new units and our own format. It is a curriculum aimed specifically at students in the algebra-based physics course.

In this article, we discuss 1) the needs, motivation, interests and learning styles of the students in the algebra-based physics course, 2) the goals of the course, 3) the curriculum and its implementation, 4) assessment, 5) open issues and 6) future developments.

**I. The student population**

The needs (both content and skills), motivation, interests and learning styles of the students in the algebra-based physics course are significantly different from those of other populations of physics students, in particular from those in the calculus-based physics course. Based on informal surveys, our experience teaching this course, and discussion with colleagues at other universities, some common characteristics of the student population are:

1) A high percentage of the students intend future careers in the health sciences (medicine, pharmacy, dentistry, physical therapy, chiropractic, etc.).

2) Most of the students are highly motivated (to make A's) and will work very hard at what they are required to do.

3) The incoming content knowledge and skill level (particularly mathematical) of the students may span a very broad range.

4) The algebra-based physics course is a freshman/sophomore level course, but most of the students are juniors or seniors. In many cases, this is because the students perceive physics as being hard and/or are afraid of physics.

5) Many of these students have done exceptionally well in school by memorizing and have not had classes that required them to

- develop concepts based on experimentation
- mathematically model experimental results
- understand how models of the physical world are developed
- solve physics problems

6) The population is about 50% female and 50% male.

7) Two semesters of algebra-based physics will be the terminal (possibly only) physics courses that they will take.

8) A minimum of about 10% of the students will take the Medical College Admissions Test (MCAT) or some other pre-professional entrance exam. (This percentage varies significantly by University.) This is the only physics course they will have before taking the exam.

9) Many of the students have poor mathematical skills at the level of algebra.

In designing our course materials, we have made the following assertions:

1) Because the needs (both content and skills), motivation, interests and learning styles of the students in the algebra-based physics course are significantly different from those of other populations of physics students, the goals of the course should be significantly different than the goals of other physics courses.

2) It is not sufficient to simply present a version of the calculus-based course, presenting the same topics in the same manner as is done in the calculus-based course, without using calculus, and possibly adding some discussion of health science applications. One must also take the needs, motivation, interests and learning styles of students in this course into account in the design of the materials. In particular, the *skills* the students should acquire and the *content* taught in the algebra-based physics course should be different than those taught in other physics courses.

**II. The goals of the course**

Defining the goals for students in this course is complex. It involves discussions which include consideration of a number of factors:

- the content and skill needs of the students for future careers

- the content and skill needs of the students for various professional school entrance exams

- the most appropriate instructional methods for this group of students, taking into account their motivations and learning styles

- the time constraints of a two semester physics sequence

- the motivations and interests of these students – the context

- the results of physics education research

When one examines these factors in detail, it becomes clear that it is not easy to design a course in which students gain a solid understanding of physics content and learn necessary scientific skills taught by appropriate teaching methods in a two semester sequence, considering the learning styles and motivations of the students. Choices have to be made *in particular* about the content and skills to be taught, if one is to achieve the goal that students retain some knowledge of physics content and the ability to use certain scientific skills in the future.

Taking the characteristics of the student population and the above factors into account, we have tried to address the following goals in our curriculum:

1) Students learn physics concepts that will be useful to them in their future careers.
   For example, students might spend more time on fluid flow, optics and waves because there are many health science applications (modeling the circulatory system, how we see and how we hear) and less time on other topics.

2) Students learn scientific skills that might be useful to them in future careers.
   For example, students learn how to develop models based on experimental results, to understand how models of the physical world are developed.

3) Students learn physics concepts that will help them pass pre-professional exams.

4) Students develop an epistemology closer to a scientist's epistemology.
   For example, students learn that understanding science includes learning about scientific process, like how to develop models based on experimental results, being able to apply concepts learned and models developed to solve problems and not just memorizing a lot of facts.

We chose to address these goals in a non-traditional format, choosing a laboratory-based environment, where students work in groups, learning about the world around them through experimentation, learning to develop both quantitative and qualitative models based on their observations and inferences. The course is also an "inquiry-based" course, in the manner of Physics by Inquiry, developed by the Physics Education Group at the University of Washington[5]. It includes Socratic dialog discussions and incorporates the results of physics education research.

### III. The curriculum and implementation

The curriculum was designed to be taught in a laboratory-based environment – no lecture – and consists of the laboratory materials, including pretests, and readings and exercises to supplement the laboratory work. It is designed to be used without a text, however, a text may be used.

The course has been taught for 8 semesters at Texas Tech University (TTU) and at Grand Valley State University (GVSU). It has been taught, while under development, to two "small" sections (24 – 30 students) of the large introductory courses at those universities each semester. Students registered for these sections on a first-come, first serve basis. The implementation was slightly different at each site. At TTU, students met for 5 contact hours per week, divided into three class sessions. At GVSU, it was a 6 contact hour class.

At each site, the students worked through the units in class, using the computers and experimental equipment. At TTU, the course was taught without a text. Students worked in groups of approximately 4, typing their groups' and individual thoughts, ideas, and answers into the computer, so all of the students' class work was on the computer for both students and instructors to access. Students could copy the materials to disk or print them out, as needed. The instructors (a faculty member and a graduate student) circulated about the room and checked the students' understanding at specific points in the material, using Socratic dialog. Homework problems were assigned weekly and were graded. Journals were used to get students thoughts on the class and allow students to express concerns. There were also exams, quizzes and a project. Homework, quiz and exam problems were designed so that students had to demonstrate their understanding of concepts, both through verbal discussion and by the use of mathematical models to solve problems, i.e. to apply their conceptual understanding in the working of quantitative problems. There was one group problem on each exam. On the group problem, students were allowed to discuss the problem with their group. Each student would hand in their own solution to the group problem, allowing a student to disagree with their group. The students were allowed to use their notes and access the computers or work experiments on the exams.

At GVSU, the course was set up in a similar manner. The students worked in groups of 2–4, depending on available equipment and content. For the first year, the materials were given to the students, unit by unit, in printed form. Subsequent to this the material was dispensed as a workbook. The students also had a standard text. The students worked through the units in class, using the computers and experimental equipment. The instructors (faculty member and undergraduate teaching assistant) circulated about the room, checking on student work. The students' written materials were periodically collected and checked. Homework problems were assigned weekly from the text with some additional problems based specifically on laboratory activities. Homework was graded. Journal entries were assigned throughout the semester. Journals were used to get students thoughts on the class and allow students to express concerns. There were also exams and quizzes. Some exams required the use of the computer based on skills obtained during this class.

All of the course materials, as taught at TTU, are on the web at http://www.phys.ttu.edu/~batcam under Courses.

IV. Assessment

The curriculum has been continually re-written based on students' performance on homework, quiz and exam questions and on the instructors' classroom observations of students' understanding. We have interviewed students on concepts related to specific topics, such as fluid dynamics, and are analyzing pretests and final exams for evidence of changes in their cognitive structure. While we believe that assessment instruments that require students to explain their reasoning are more appropriate indicators of success at our goals, we have also administered the Force Concept Inventory (FCI)[6] and Maryland Physics Expectation Survey (MPEX)[7] as pre- and post- tests.

We have evidence of a shift in students' epistemology from viewing physics class as something that has to be endured (which involves a focus on the process of getting a grade), to something more than tolerable, even enjoyable, which involves a focus on learning about the world around them through experiments they do themselves in class, on the process of science. We have also watched them move from a focus on memorization, rote problem solving algorithms and getting the right answer, to a focus on the process of problem solving as the process of applying models they have developed that are consistent with their experimentation to other situations. The evidence of this can be found in interview transcripts and the students' solutions to examination problems. A detailed analysis is in progress.

The FCI scores at TTU have an absolute gain of about 30% and an averaged normalized gain of about 0.4, falling in the "interactive engagement region" and in the "medium–g" range as defined by Hake[8] and significantly better than the FCI scores in the TTU classes taught traditionally. TTU MPEX scores showed a significant increase in the "coherence cluster", students seeing physics as a "coherent, consistent structure" and no decrease in any of the clusters in the survey, as often seen in physics courses across the country.

**V. Course implementation issues**

There are a number of issues that arise in the development of a course specifically for the algebra-based physics class. In designing a curriculum, we have had to make choices on content, coverage, skills taught, learning environment, teaching style, etc. However, curricula are not (should not be) static, and some of these issues are things that faculty and administrators at many universities need to address. We open these for discussion, and hope for a national discussion of these issues, as related to the algebra-based physics course.

**A. Content coverage**

We have struggled extensively with the issue of content coverage. The algebra-based physics course is often taught in two semesters and the coverage is expected to include content through modern physics. This is a much wider breadth of content, than expected in most calculus-based physics courses. Supposedly, the topics are to be covered in less depth. This does not make sense to us, for number of reasons. The students enter this course with similar (incorrect) pre-conceptions as the students in the calculus-based physics course and one would expect it to take at least the same amount of time to develop a conceptual understanding consistent with a scientists'. (The pre-FCI scores for algebra-based physics students are usually of the order 0.25, about the same as pre-FCI scores for high school students, while the average pre-FCI score for calculus-based physics students is often in the 0.4 to 0.5 range for most universities.) One might expect that it would take more time to develop a strong conceptual understanding in this class than for students in the calculus-based physics class, not less. One might also argue that it would benefit the students more to develop a solid conceptual understanding of the topics covered, than to have a very shallow exposure to many topics, and not have developed a solid understanding of any of them.

The students in the algebra-based physics class often have very weak math skills and often have done very little problem solving. One has to make decisions as to whether to address these skills as part of the curriculum. If time is taken within the curriculum to address these skills, then it is not possible to cover as many topics. In addition, if time is to be spent on understanding the process of science, on learning how to develop models based on experimental results, it is not always possible to cover as many topics.

Another issue that arises is the question of whether a greater focus should be placed on topics that will be useful to students in their future careers. This might mean, that more time might be spent on topics such as fluids, optics and radioactivity than on kinematics and forces, which is not always the case in a traditional curriculum. Decisions have to be made about which topics to cover and the amount of time to spend on each topic.

There is one argument counter to the above arguments that may suggest that a breadth of coverage is better, and that is that most of the pre-professional exams, such as the MCAT, require knowledge of physics through modern physics in the physical science section of the exam. This complicates decisions of content coverage, and implies that it is important to cover topics through modern physics in this course. However, another consideration is that the MCAT is also changing, requiring more qualitative understanding of concepts and focusing more on skills other than memorization, which supports the coverage of less topics in more depth.

**B. Institutional issues**

The design and implementation of curricula are often complicated by institutional and other issues. Many physics departments, including some of the departments in which these curricula are being developed, are used to:

    1) a traditional mode of instruction,

    2) have not considered the non-content goals of the course listed above to be important, but only the amount of content to be covered,

    3) have not paid attention to research on students' understanding,

    4) are often concerned with maintaining a high student/professor ratio (this is often an administrative issue) and

    5) are often more concerned with graduate students and physics majors than with the introductory algebra-based physics course.

These issues often impose major barriers to the implementation of the type of course we have developed, even though individual faculty members are interested in implementing the course and have found ways to surmount institutional barriers. It is possible, for example, to teach large introductory classes in an inquiry-based, laboratory-based environment with approximately the same number of faculty and graduate TA's presently assigned to the course, by effectively using their time and using undergraduate teaching assistants. Examples of this are Studio Physics[9] classes and Student-Centered Activities for Large Enrollment Undergraduate Programs**,** SCALE-UP[10] classes, taught to students in calculus-based physics classes. The same could be done for algebra-based physics classes. The limitation often arises more with equipment and laboratory classroom space, than with the instructor to student ratio. It simply takes an institutional commitment to make it happen.

We hope that faculty and administrators will open discussion on implementing this type of teaching environment as a possibility for students in large introductory courses at more institutions.

**VI. Future plans**

At TTU, the course is being expanded to be taught to pre-service middle school mathematics teachers and as the honors section of the algebra-based physics course, in addition to a special section of the large introductory physics class. We are also exploring the possibility of another adaptation of the materials to be taught at the high school level and as a ninth grade Integrated Physics and Chemistry (IPC) course.

The main facilitator at GVSU has moved to a different institution. At present, two sections of the class are still being taught in the format mentioned earlier.

We hope to continue and expand assessment of the course and continue offering workshops about the curriculum to high school and university instructors at American Association of Physics Teachers (AAPT) meetings, depending on funding.

We have been contacted by faculty at other institutions about adapting parts of the materials or offering them in a SCALE-UP environment. We hope to continue to work with those institutions.

**VII. Conclusion**

We have developed a curriculum specifically for the introductory algebra-based physics course, taking into account the needs, backgrounds, learning styles and career goals of the students in that class. The course is designed to be taught in an inquiry-based, laboratory-based environment, however parts of the materials can be used in other settings. We have modified and adapted parts of existing materials and integrated them with our own new units and our own format, creating a course aimed specifically at these students. We would like to thank the National Science Foundation for funding Adaptation and

Implementation grants, because we think this type of grant is crucial to helping institutions adapt materials from other environments to their own university and, in our case, adapting materials developed for other student populations to our population. We also hope that faculty and administrators will open discussions about content coverage and institutional issues as related to the algebra-based physics course at many universities and that we can have a national discussion of the issues.

This article is based upon work supported by the National Science Foundation under Grant Nos. CCLI #9981031 and CCLI-EMD #0088780. Any opinions, findings and conclusions or recommendations expressed are those of the author(s) and do not necessarily reflect the views of the National Science Foundation (NSF).